\documentclass{JHEP3}

\usepackage{epsfig}
\usepackage{cite}

\newcommand{\eps}{\epsilon}  
\newcommand{\alps}{\alpha_{\mathrm{s}}}
\newcommand{\zi}{\tilde z_i}
\newcommand{\zj}{\tilde z_j}

\newcommand{\yijk}{y_{ij,k}}
\newcommand{\vijk}{v_{ij,k}}
\newcommand{\viji}{v_{ij,i}}                                                    
\newcommand{\tvijk}{\tilde v_{ij,k}}  
\newcommand{\bV}       {{\bf V}}
\newcommand{\CF}{C_{\mathrm{F}}}
\newcommand{\CA}{C_{\mathrm{A}}}
\newcommand{\TR}{T_{\mathrm{R}}}  
\newcommand{\cD}{{\cal D}}
\newcommand{\tpij}{\widetilde p_{ij}}
\newcommand{\tpk}{\widetilde p_k}
\newcommand{\xija}{x_{ij,a}}
\newcommand{\xiab}{x_{i,ab}}
\newcommand{\tpa}{\tilde p_a}
\def\bom#1{{\mbox{\boldmath $#1$}}}
\newcommand{\tpai}{\tilde p_{ai}}
\newcommand{\tpj}{\tilde p_j}

\newcommand{\dlpl}{\delta_{\lambda'\lambda}}
\newcommand{\dlli}{\delta_{\lambda\lambda_i}}
\newcommand{\dllj}{\delta_{\lambda\lambda_j}}
\newcommand{\dlla}{\delta_{\lambda\lambda_a}}
\newcommand{\dlilj}{\delta_{\lambda_i\lambda_j}}
\newcommand{\dlali}{\delta_{\lambda_a\lambda_i}}
\newcommand{\dplli}{\delta'_{\lambda\lambda_i}}

\newcommand{\dplla}{\delta'_{\lambda\lambda_a}}
\newcommand{\dplilj}{\delta'_{\lambda_i\lambda_j}}
\newcommand{\dplali}{\delta'_{\lambda_a\lambda_i}}

\newcommand{\dllQ}{\delta_{\lambda\lambda_Q}}
\newcommand{\dllQb}{\delta_{\lambda\lambda_{\bar Q}}}
\newcommand{\dllg}{\delta_{\lambda\lambda_g}}
\newcommand{\dlQlg}{\delta_{\lambda_Q\lambda_g}}

\newcommand{\dpllQ}{\delta'_{\lambda\lambda_Q}}

\newcommand{\dplQlg}{\delta'_{\lambda_Q\lambda_g}}
\newcommand{\dplQlQb}{\delta'_{\lambda_Q\lambda_{\bar Q}}}

\def\asymp#1%
{\mathrel{\raisebox{-.4em}{$\widetilde{\scriptstyle #1}$}}}

\title{\bf Polarizing the Dipoles }

\author{M. Czakon$^{a,b}$, C. G. Papadopoulos$^{a}$ and M. Worek$^{b,c}$
  \\ \\
  $^a$~{Institute of Nuclear Physics, NCSR Demokritos,
    GR-15310 Athens, Greece}
  \\
  $^b$~{Fachbereich C, Bergische Universit\"at Wuppertal, D-42097, Wuppertal,
    Germany}
  \\
  $^c$~{Department of Field Theory and Particle Physics,
    Institute of Physics, \\
    University of Silesia, Uniwersytecka 4, PL-40007 Katowice,
    Poland} }

\abstract{We extend the massless dipole formalism of Catani and
  Seymour, as well as its massive version as developed by Catani,
  Dittmaier, Seymour and Trocsanyi, to arbitrary helicity eigenstates
  of the external partons. We modify the real radiation subtraction
  terms only, the primary aim being an improved efficiency of the
  numerical Monte Carlo integration of this contribution as part of a complete
  next-to-leading order calculation. In consequence, our extension is
  only applicable to unpolarized scattering. Upon summation over the
  helicities of the emitter pairs, our formulae trivially reduce to
  their original form. We implement our extension within the framework
  of \textsc{Helac-Phegas}, and give some examples of results
  pertinent to recent studies of backgrounds for the LHC. The code is {\it
  publicly} available. Since the integrated dipole contributions do not
  require any modifications, we do not discuss them, but they are
  implemented in the software.}

\preprint{WUB/09-03}

\begin{document}

%%%%%%%%%%%%%%%%%%%%%%%%%%%%%%%%%%%%%%%%%%%%%%%%%%%%%%%%%%%%%%%%%%%%%%%%%%%%%%%%

\section{Introduction}
\label{sec:introduction}

At present, the necessity of next-to-leading order (NLO) calculations
of QCD backgrounds for the Large Hadron Collider (LHC) is
unquestionable. Much effort has been put into this problem, but until
only recently, it seemed that the task is so huge that theory will
stay behind the needs of experimentalists for quite some time. Whereas
impressive calculations have been done with traditional methods based
on Feynman diagrams
\cite{Han:1991ia,deFlorian:1999tp,Beenakker:2002nc,
  DelDuca:2003uz,Figy:2003nv,Nagy:2003tz,Oleari:2003tc,Jager:2006zc,
  Jager:2006cp,
  Campbell:2006xx,Bozzi:2007ur,Dittmaier:2007wz,Ciccolini:2007jr,
  Dittmaier:2007th,Campbell:2007ev,Ciccolini:2007ec,Hankele:2007sb,
  Bredenstein:2008zb,Campanario:2008yg,Dittmaier:2008uj,Bredenstein:2009aj},  
  it is the new unitarity based methods \cite{Bern:1994zx,Bern:1994cg,
  Witten:2003nn,Britto:2004nc,Brandhuber:2005jw,Britto:2006sj,Bern:2007dw}
that  provide some hope for accelerated progress. By now, there are
three major  groups with advanced software for virtual corrections
\cite{Berger:2006ci,
  Berger:2006vq,Berger:2008sj,Berger:2008ag,Ellis:2006ss,Giele:2008bc,
  Ossola:2006us,Ossola:2007bb,Ossola:2007ax,Ossola:2008xq,Mastrolia:2008jb,
  Draggiotis:2009yb,vanHameren:2009dr}, closely trailed by independent
efforts \cite{Lazopoulos:2008ex}. Moreover, a full automate based on
traditional methods is being built 
\cite{Binoth:1999sp,Binoth:2005ff, Binoth:2006hk,Binoth:2008gx}. In
any case, first successes have been recorded in
\cite{Lazopoulos:2007ix,Lazopoulos:2007bv,Binoth:2008kt,Ellis:2008qc,
  Ellis:2009zw,Berger:2009zg} and look very promising.

Any NLO calculation consists of two parts, which are separately
infrared (soft/collinear) divergent, the virtual corrections and real
radiation. In order to allow for Monte Carlo simulations, two general
classes of approaches have been devised, namely phase space slicing
\cite{Fabricius:1981sx,Kramer:1986mc,Baer:1989jg,Giele:1991vf,Giele:1993dj,
  Harris:2001sx} and subtraction \cite{Ellis:1989vm,
  Kunszt:1992tn,Catani:1996jh,Catani:1996vz,Nagy:1996bz,Frixione:1997np,
  Catani:2002hc}. Currently, the latter approach seems to have proven
its superiority. Its most accepted version, dipole subtraction, has
been presented in the massless case in \cite{Catani:1996vz} and later
for massive partons in \cite{Catani:2002hc} (see also
\cite{Phaf:2001gc}). There are a few 
completely automated implementations of this method, which have been
presented in the literature \cite{Gleisberg:2007md,Seymour:2008mu,
  Hasegawa:2008ae,Frederix:2008hu}, but only one is available for
public use \cite{Frederix:2008hu}. Moreover, it has often been
criticized that the large number of terms in the dipole subtraction
formalism is a practical problem in realistic calculations, since it
leads to a high computational costs. With the present publication we
want to remedy both problems.

The basic idea is to allow for the same optimizations as those used in
leading order simulations. The most important of these, besides phase
space optimization, is the replacement of exact summation over
external state polarizations by a probabilistic approach. In order not
to ruin the Monte Carlo error estimates, the treatment must be
consistent between the real emission contribution and the appropriate
subtraction. This is the main subject of the present
publication. Note that partial extensions of the dipole subtraction
formalism to polarized states have been discussed before in
\cite{Dittmaier:1999mb,Dittmaier:2008md}, but they were only concerned
with scattering of polarized fermions or photons. 
The formulae from these studies
are consistent with ours, when summed over the polarization of the
gluon (or photon as discussed by these authors). 
Besides giving the appropriate formulae, we implement
them within the framework of a fully fledged Monte Carlo generator,
\textsc{Helac-Phegas} \cite{Kanaki:2000ey,Papadopoulos:2000tt,
Cafarella:2007pc}, which has, on its own, already been extensively
used and tested in phenomenological studies (see for example
\cite{Gleisberg:2003bi,Papadopoulos:2005ky,Alwall:2007fs,Englert:2008tn}). We
demonstrate the potential of the software by performing some realistic
simulations, in particular of the process $gg \rightarrow t {\bar t} b
{\bar b} g$\footnote{While this publication was being prepared, a complete
study of the full hadronic process has been published
\cite{Bredenstein:2009aj}.}. We also argue that the inclusion of the
subtraction terms does not increase the total evaluation time per
phase space point by large factors. In fact, the actual source of the
substantially longer evaluation times in comparison to a leading order
evaluation is the more complicated phase space, requiring orders of
magnitude more accepted points to reach the same accuracy. In
consequence, one can either further improve the evaluation time per
phase space point by using color sampling for example (trivial in our
approach), or concentrate on a better description of the phase
space. We leave both tasks to future studies.

The paper is organized as follows. In the next section, we discuss the
r\^ole and treatment of polarization in Monte Carlo generators, which
leads us to the motivation for the present study. Subsequently, we
describe the behavior of cross sections in soft and collinear limits,
when polarized partons are present. Section \ref{sec:dipole} contains
our main results, namely the dipole subtraction formulae  for
polarized external partons. Section \ref{sec:implementation}, on the
other hand, presents a few details of our implementation within 
\textsc{Helac-Phegas}, as well as some realistic simulation results. We
conclude in Section \ref{sec:conclusions}.

%%%%%%%%%%%%%%%%%%%%%%%%%%%%%%%%%%%%%%%%%%%%%%%%%%%%%%%%%%%%%%%%%%%%%%%%%%%%%%%%

\section{Polarization in Monte Carlo simulations}
\label{sec:polarization}

Most practical problems, which are solved with Monte Carlo simulations
involve unpolarized particles. In consequence, it is necessary to
sum/average over the spin of the incoming and outgoing states. This
increases the computational complexity of a calculation by a factor, which
can, in principle, amount to $2^{n_2} 3^{n_3}$, where $n_2$ and $n_3$ are
the numbers of particles with 2 and 3 polarization states
respectively. There are usually some symmetries, like the
chiral symmetry in the massless fermion approximation, or
supersymmetry in the pure gluon case, which reduce the number of
degrees of freedom. Moreover, with modern recursive methods for tree
level matrix element evaluation it is possible to reuse parts of a
result to speed up the summation. Nevertheless, deterministic, exact
methods have an inherent slow down factor, which cannot be completely
removed.

Since the phase space integration is already done with probabilistic
methods, and the polarization sum is not coherent, it is clearly
desirable to replace this sum by some kind of random sampling as
well. The approach, which is most often used is to sample over
helicity. The main disadvantage here is that different helicity
configurations contribute very differently to the final result. In
fact, several orders of magnitude between contributions are usually
observed. If we use a flat distribution to pick a helicity
configuration, the final variance of a result will be increased by a
substantial factor, not to mention the increase in the number of
generated phase space points to obtain a given error estimate. The
situation can be improved by using stratified sampling, which is
usually also used for phase space optimization. One of the possible
algorithms, which we implemented in our software, has been presented
in \cite{Kleiss:1994qy} (more details will follow in
Section~\ref{sec:implementation}).

Another approach, which is substantially easier to implement has been
proposed in \cite{Draggiotis:1998gr,Draggiotis:2002hm}. The idea is to
replace summation over helicity by integration over a phase. For
example, a gluon polarization state can be written as
\begin{equation}
\label{eq:pol}
\eps_\mu(k,\phi) = e^{i \phi}\eps_\mu(k,+) + e^{-i \phi}\eps_\mu(k,-)
\; ,
\end{equation}
where $\eps_\mu(k,\pm)$ are helicity eigenstates. The sum over the
helicity of this gluon fulfills the following identity
\begin{equation}
\sum_{\lambda} |M_\lambda|^2 = \frac{1}{2\pi} \int_0^{2\pi} d \phi |M_\phi|^2
\; .
\end{equation}
Notice that the range of integration could have been reduced to
$[0,\pi]$ with the same result. We keep $2\pi$ as the upper bound in order
to accommodate the third degree of freedom of massive gauge bosons,
which is then added in Eq.~\ref{eq:pol} without any phase
factor. Obviously, we do not expect large differences in the values of
$|M_\phi|^2$, as function of $\phi$, since for every value of $\phi$
we have both helicities contributing. In consequence, a flat
distribution in the Monte Carlo sampling should provide satisfactory
results, which has been confirmed on specific examples in
\cite{Draggiotis:1998gr,Draggiotis:2002hm} and also in our studies.

At this point, we should decide which approach to chose for dipole
subtraction. As we will show in the next Section, helicity eigenstates
provide particularly simple formulae for this problem, which are only
minor modifications of the original formalism. Therefore, we trade the
simplicity of the implementation of the Monte Carlo integration over
general polarizations with a phase, for simple dipole subtraction
formulae. Practice also shows that whereas helicity sampling is
superior for a small number of helicity configurations, with many
particles general polarizations with phases start to dominate. The
reason is that at some point the algorithm of helicity sampling cannot
find the optimal distribution. In our case, where the number of
partons is relatively low, since we are always thinking of a
next-to-leading order calculation, which is bound in complexity by our
ability to compute virtual corrections, the large number of final
states comes from electroweak decays into colorless states. We found
it optimal to use a hybrid model, where partons have definite
helicities and remaining particles have general polarizations with
phases.

%%%%%%%%%%%%%%%%%%%%%%%%%%%%%%%%%%%%%%%%%%%%%%%%%%%%%%%%%%%%%%%%%%%%%%%%%%%%%%%%

\section{Soft and collinear limits for polarized partons}
\label{sec:soft}

In view of the considerations of the previous Section, we have two
possibilities to treat the polarization of partons. Indeed, we can
either use arbitrary polarization vectors, or helicity
eigenstates. Let us first show that, when using the latter the soft
limit is particularly simple.

It is well known that the exchange of a soft gluon between two partons
(quarks or gluons) can be approximated by eikonal currents
\begin{equation}
|M_{m+1}|^2 \sim {}_m\langle\dots|\bom J^{\mu,a\;\dagger} \bom
J^{a}_{\mu}|\dots\rangle_m \; ,
\end{equation}
where we have omitted irrelevant constants and the current $\bom J$
is given by
\begin{equation}
\label{eq:current}
\bom J^{\mu,a} = \sum_i \bom T^a_i \frac{p_i^\mu}{p_i k} \; ,
\end{equation}
with $k$ the momentum of the soft gluon and $p_i$ the momentum of a
hard parton. The color charge operators $\bom T^a$ are defined as in
  \cite{Catani:1996vz}, {\it i.e.} for two given color space vectors
$|a_1,\dots,a_m\rangle$ and $|b_1,\dots,b_n\rangle$, we have
\begin{equation}
\langle a_1,\dots,a_i,\dots,a_m|\bom
T^c_i|b_1,\dots,b_i,\dots,b_m\rangle = \delta_{a_1b_1}\dots
T^c_{a_ib_i}\dots\delta_{a_mb_m} \; ,
\end{equation}
where $T^a_{bc}=i f_{bac}$ (color charge matrix in the adjoint
representation) if parton $i$ is a gluon,
$T^a_{\alpha\beta}=t^a_{\alpha\beta}$ (color charge matrix in the
fundamental representation) if parton $i$ is an outgoing quark, and
$T^a_{\alpha\beta}=-t^a_{\beta\alpha}$ if parton $i$ is an outgoing
anti-quark. The charges of in-going partons are defined by
crossing. With this definition
\begin{equation}
\sum_i \bom T^a_i = 0 \; ,
\end{equation}
which means that no signs are needed in Eq.~\ref{eq:current}, and the
current is both transverse and self-conjugate
\begin{equation}
k_\mu \bom J^{\mu,a} = 0 \; , \quad \quad \bom J^{\mu,a \; \dagger} =
\bom J^{\mu,a} \; .
\end{equation}
While it is clear that the eikonal approximation, and thus also the
soft limit, is independent of the polarization of the hard partons, if
the soft gluon is polarized on the other hand, we have
\begin{equation}
\bom J^{\mu,a\;\dagger} \bom J^{a}_{\mu} \; \longrightarrow \;
-\bom J^{\mu,a\;\dagger} \bom J^{\nu,a} \eps_\mu(k,\lambda)
\eps^{*}_\nu(k,\lambda) \; .
\end{equation}
Crucially, for helicity eigenstates
\begin{equation}
\label{eq:helicity}
\eps^{*}_\mu(k,+) = e^{i \psi} \eps_\mu(k,-) \; ,
\end{equation}
where $\psi$ is some phase, which can be freely chosen. With this
relation and the properties of the eikonal current, it is easy to show
that
\begin{equation}
\bom J^{\mu,a\;\dagger} \bom J^{\nu,a} \eps_\mu(k,+) \eps^{*}_\nu(k,+)
=  \bom J^{\mu,a\;\dagger} \bom J^{\nu,a} \eps_\mu(k,-)
\eps^{*}_\nu(k,-) = \frac{1}{2} \bom J^{\mu,a\;\dagger} \bom J^a_\mu \; .
\end{equation}
Thus, we have shown that the soft limit is independent of the helicity
of the soft gluon. This is important, because this means that we can
introduce helicity into the unpolarized dipole subtraction formulae
without explicit reference to the polarization vectors of the soft
gluons. Unfortunately, for general polarization vectors, such as those
defined in Section~\ref{sec:polarization}, there is no relation of the
type of Eq.~\ref{eq:helicity}. In consequence, if we wanted to use general
polarization vectors, a less trivial modification of the dipoles would
be needed.

Let us now turn to the collinear limit. It is in principle possible to
work directly with amplitudes, however due to the treatment of the
soft limit, the original dipole subtraction formalism has been
formulated for squares of matrix elements. Here we shall proceed
similarly.

We consider the quasi-collinear limit for massive partons and the
corresponding true collinear limit in the massless case. For a pair
$\{i,j\}$ of outgoing partons, which become collinear, we assume the
following momentum parameterization
\begin{eqnarray}
p_i^\mu &=& z p^\mu + k_\perp^\mu 
- \frac{k_\perp^2+ z^2 m_{ij}^2- m_i^2}{z}\frac{n^\mu}{2pn} \; , \\
p_j^\mu &=& (1-z) p^\mu - k_\perp^\mu 
- \frac{k_\perp^2+(1-z)^2m_{ij}^2 - m_j^2}{1-z}\frac{n^\mu}{2pn} \; ,
\end{eqnarray}
where $p_i^2 = m_i^2$, $p_j^2 = m_j^2$ and $p^2 = m_{ij}^2$, $n$ is a
light-like auxiliary vector, and $k_\perp$ is the transverse component
orthogonal to both $p$ and $n$, which parameterizes the collinear
limit. The parton of mass $m_{ij}$ is the virtual particle, which
splits into $i$ and $j$. Its nature is uniquely determined in
QCD. For example, if $i$ is a quark and $j$ is an anti-quark, then
$m_{ij} = 0$ corresponds to a virtual gluon. As in
\cite{Catani:2002hc}, we define the limit by a uniform rescaling
\begin{equation}
k_\perp \to \alpha k_\perp \; , \quad m_i\to \alpha m_i \; , 
\quad m_j\to \alpha m_j \; , \quad m_{ij}\to \alpha m_{ij} \; ,
\end{equation}
with $\alpha \rightarrow 0$. The matrix element behaves in this limit
as
\begin{eqnarray}
\lefteqn{
{}_{m+1}\langle \dots,\{p_i,\lambda_i\},\dots,\{p_j,\lambda_j\},\dots||
\dots,\{p_i,\lambda_i\},\dots,\{p_j,\lambda_j\},\dots\rangle_{m+1}
\;\; \asymp{\alpha\to 0} 
}
\\ \nonumber
&& \quad \quad
\frac{1}{\alpha^2} \frac{8\pi\alps}{(p_i+p_j)^2-m_{ij}^2} \:
{}_m\langle \dots,\{p,\lambda'\},\dots|
\hat P_{\widetilde{ij},i}(z,k_\perp,\{m\},\{\lambda\})
|\dots,\{p,\lambda\},\dots\rangle_m \; ,
\end{eqnarray}
where $\hat P$ are generalized Altarelli-Parisi kernels (here in four
dimensions), $\{m\}$ is the set of masses, and $\{\lambda\}$ is the set of
helicities, whereas $\widetilde{ij}$ is the emitter parton (we will
call the original pair, the emitter pair).

Whereas the unpolarized massive case of $\hat P$ has been presented in
\cite{Catani:2002hc}, the polarized massless case can be read to a
large extent already from \cite{Altarelli:1977zs}. Here we present the
formulae, which contain all the information
\begin{eqnarray}
\lefteqn{
\langle\lambda'|P_{QQ}(z,k_\perp,m_Q,\lambda_Q,\lambda_g)|\lambda\rangle =
}
\\ \nonumber && \quad \quad
\CF \left\{ \frac{\dllQ(z^2+\dlQlg(1-z^2))}{1-z} -
\dllg\dplQlg \frac{m_Q^2}{p_Q p_g} \right.
\\ \nonumber && \quad \quad \quad \quad \left.
-\left[ \left(\dllQ\dlQlg-\dpllQ\dplQlg\right)\frac{1}{z}
+\dplQlg(\dllQ-\dllg)z \right] \frac{m_Q^2}{2p_Q p_g}
\right\} \dlpl \; ;
\\ \nonumber \\
\lefteqn{
\langle\lambda'|P_{gQ}(z,k_\perp,m_Q,\lambda_Q,\lambda_{\bar Q})|
\lambda\rangle =
}
\\ \nonumber && \quad \quad
\TR \left\{
\frac{\dlpl\dplQlQb}{2}-\frac{2\dplQlQb
\left(k_\perp \cdot \eps(p,\lambda')\right)^{*}
\left(k_\perp \cdot \eps(p,\lambda)\right)}{(p_Q+p_{\bar Q})^2} \right.
\\ \nonumber && \quad \quad \quad \quad \left.
+ \dlpl \left[ \dplQlQb \left( \dllQ z + \dllQb (1-z) - \frac{1}{2}
  \right)
\right. \right. \\ \nonumber && \quad\quad\quad\quad\quad \left.\left.
+\left( \left(\dllQ-\dplQlQb\right) \frac{1}{z}
+ \left(\dllQb-\dplQlQb\right) \frac{1}{1-z}
\right) \frac{m_Q^2}{(p_Q+p_{\bar Q})^2} \right]
\right\} \; ;
\\ \nonumber \\
\lefteqn{
\langle\lambda'|P_{gg}(z,k_\perp,\lambda_i,\lambda_j)|
\lambda\rangle =
}
\\ \nonumber && \quad \quad
\CA \left\{
\dlpl \left( \frac{\dlli}{1-z} +\frac{\dllj}{z}-2\dplilj \right)
-2\dplilj z(1-z) \frac{
\left(k_\perp \cdot \eps(p,\lambda')\right)^{*}
\left(k_\perp \cdot \eps(p,\lambda)\right)}{k_\perp^2} \right.
\\ \nonumber && \quad \quad \quad \quad \left.
+ \dlpl \left[ \dlli - \dllj \right] (1-2z)
\right\} \; ;
\end{eqnarray}
where $\dplilj=1-\dlilj$. Notice that the contents of the square
brackets vanish upon summation over the helicities of the emitter
partons. Because the soft limit remains untouched with helicity
eigenstates, the above formulae alone can be used to modify the dipole
subtraction terms. An important difference to previous studies is that
we do not use open Lorentz indices, but contract directly with
polarization vectors. This is allowed by the transversality of the
$k_\perp$ vector (and its counterparts in the final dipole subtraction
formulae) and of the matrix element. Moreover, we used the phase
conventions for the polarization vectors as given in
\cite{Kanaki:2000ey}, which are also consistent with those of
\textsc{MadGraph} \cite{Maltoni:2002qb}. Any difference in 
conventions should be relatively easy to compensate.

Notice finally, that the initial state splittings can be derived from
the above formulae by crossing.

%%%%%%%%%%%%%%%%%%%%%%%%%%%%%%%%%%%%%%%%%%%%%%%%%%%%%%%%%%%%%%%%%%%%%%%%%%%%%%%%

\section{Dipole subtraction with helicity eigenstates}
\label{sec:dipole}

The dipole subtraction formalism has been described to great extent in
\cite{Catani:1996vz} and \cite{Catani:2002hc}. We do not repeat this
discussion and assume that the reader is familiar with the main
concepts. On the other hand, we give all the necessary formulae for a
complete implementation in a numerical program, {\it i.e.} not only
the modified splitting kernels, but also the momentum remappings.

The starting point of our exposition is the subtracted real radiation
contribution to a next-to-leading order cross section
\begin{equation}
\label{eq:main}
\int d \Phi \sum (|M_{m+1}|^2-\cD) \; ,
\end{equation}
where the sum runs over polarizations and colors, and an average over
the initial state colors and polarizations is understood together with
a symmetry factor for the final states. We have also omitted the jet
functions in the phase space integration $d \Phi$, since they are
irrelevant at the moment. In this somewhat schematic expression, the
dipole contribution can be decomposed as
\begin{equation}
\label{eq:decomposition}
\cD =
\sum_{\{i,j\} }\sum_{k\ne i,j} \cD_{ij,k} +
\sum_{\{i,j\}}\sum_a \cD_{ij}^a +
\sum_{a,i}\sum_{j\ne i} \cD_j^{ai} +
\sum_{a,i}\sum_{b\ne a} \cD^{ai,b} \; ,
\end{equation}
where $i,j,k$ denote final states, whereas $a,b$ initial states. A
pair of indices corresponds to the emitter pair and a single index
specifies the spectator. The fact that the sum in Eq.~\ref{eq:main}
runs over the difference of the real emission matrix element squared
and the dipole subtraction contribution is not a coincidence of
course, since we want to have a cancellation of divergences for each
helicity configuration. The formulae that we present in the following
guarantee that a contribution corresponding to a given helicity
configuration of the partons is finite in soft and collinear
limits. In principle, we could also give the formulae in a form, in
which the finiteness would extend to a given color flow, but we
refrain from this in the present work.

Clearly, in each of the dipoles in Eq.~\ref{eq:decomposition} all the
polarizations, but those of the emitter pair, must be taken over from the
matrix element $M_{m+1}$. Since the sums run over all partons and our
formulae are only valid for helicity eigenstates, we have to require
that all partons be in helicity eigenstates. On the other hand, there
is no restriction on the polarization states of the remaining
particles.

Finally, let us stress that since we are only interested in the
subtraction for the real emission, we work exclusively in four
space-time dimensions.

%%%%%%%%%%%%%%%%%%%%%%%%%%%%%%%%%%%%%%%%%%%%%%%%%%%%%%%%%%%%%%%%%%%%%%%%%%%%%%%%

\subsection{Final state emitter and final state spectator}

The single dipole contribution is
\begin{eqnarray}
&&  \!\!\!\!
\cD_{ij,k}(\dots,\{p_i,\lambda_i\},\dots,\{p_j,\lambda_j\},\dots,
\{p_k,\lambda_k\},\dots) =
-\frac{1}{(p_i+p_j)^2-m_{ij}^2} \\ \nonumber
&& \quad \quad \quad \times \sum_{\lambda',\lambda}
\langle\dots,\{\tpij,\lambda'\},\dots,\{\tpk,\lambda_k\},\dots|
\frac{{\bom T}_k\cdot{\bom T}_{ij}}{{\bom T}_{ij}^2} \bV_{ij,k}
|\dots,\{\tpij,\lambda\},\dots,\{\tpk,\lambda_k\},\dots\rangle \; ,
\end{eqnarray}
where the momentum remapping  is given by ($Q=p_i+p_j+p_k$)
\begin{equation}
\tpk^\mu =
\frac{\sqrt{\lambda(Q^2,m_{ij}^2,m_k^2)}}{\sqrt{\lambda(Q^2,(p_i+p_j)^2,m_k^2)}}
\left( p_k^\mu-\frac{Q p_k}{Q^2}Q^\mu \right)
+\frac{Q^2+m_k^2-m_{ij}^2}{2Q^2} Q^\mu \; ,
\end{equation}
\begin{equation}
\tpij^\mu = Q^\mu-\tpk^\mu \; ,
\end{equation}
with $\lambda$ the K\"allen function
\begin{equation}
\label{eq:kaellen}
\lambda(x,y,z) = x^2+y^2+z^2-2xy-2xz-2yz \; .
\end{equation}
There are three cases to consider
\begin{itemize}
\item
{\boldmath $Q\to Q(p_i,\lambda_i)+g(p_j,\lambda_j)$}, with
$m_i=m_{ij}=m_Q$, $m_j=0$ and
\begin{eqnarray}
\label{eq:ff1}
\langle\lambda'|\bV_{Qg,k}|\lambda\rangle &=&
8\pi\alps\CF \\ \nonumber &\times&
\left\{
\frac{\dlli}{1-\zi(1-\yijk)}
-\dplilj \frac{\tvijk}{\vijk} \left[ \dlli (1+\zi) + \dllj \frac{m_Q^2}{p_i p_j}
\right] \right. \\ \nonumber 
&& \left. - \frac{\tvijk}{\vijk} \left[
(\dlli\dlilj-\dplli\dplilj) \frac{1}{\zi} + \dplilj(\dlli-\dllj) \zi\right]
\frac{m_Q^2}{2p_ip_j}
\right\}
\dlpl \; ;
\end{eqnarray}
\item
{\boldmath $g\to Q(p_i,\lambda_i) + \bar Q(p_j,\lambda_j)$}, with 
$m_i=m_j=m_Q$, $m_{ij}=0$ and
\begin{eqnarray}
\label{eq:ff2}
\langle\lambda'|\bV_{Q\bar Q,k}|\lambda\rangle &=& 
8\pi\alps\TR \\ \nonumber &\times&
\frac{1}{\vijk} 
\left\{
\dlpl\left[ \frac{\dplilj}{2} - \frac{\kappa}{2}
\left(z_+z_- -\frac{m_Q^2}{(p_i+p_j)^2}\right) \right] \right. \\ \nonumber
&& \left. -\frac{2\dplilj}{(p_i+p_j)^2}
\left[\zi^{(m)}p_i^\mu-\zj^{(m)}p_j^\mu\right]
\left[\zi^{(m)}p_i^\nu-\zj^{(m)}p_j^\nu\right]
\eps^{*}_\mu(\tpij,\lambda')\eps_\nu(\tpij,\lambda) \right. \\ \nonumber
&& + \dlpl \left. \left[ \dplilj\left(\dlli \zi+\dllj \zj-\frac{1}{2}\right)
\right. \right. \\ \nonumber && \quad \quad \quad \left. \left.
+\left( (\dlli-\dplilj) \frac{1}{\zi}+(\dllj-\dplilj)\frac{1}{\zj}\right)
\frac{m_Q^2}{(p_i+p_j)^2} \right] 
\right\} \; ;
\end{eqnarray}
\item
{\boldmath $g\to g(p_i,\lambda_i) +g(p_j,\lambda_j)$}, with
$m_i=m_j=m_{ij}=0$ and
\begin{eqnarray}
\label{eq:ff3}
\langle\lambda'|\bV_{gg,k}|\lambda\rangle  &=&
8\pi\alps\CA \\ \nonumber &\times&
\left\{ \dlpl \left[\frac{\dlli}{1-\zi(1-\yijk)}
+\frac{\dllj}{1-\zj(1-\yijk)}-\frac{2\dplilj-\kappa z_+z_-/2}{\vijk}
\right] \right.\\ \nonumber
&& \left. +\frac{\dplilj}{\vijk}\frac{1}{p_i p_j}
\Big[\zi^{(m)}p_i^\mu-\zj^{(m)}p_j^\mu\Big]
\Big[\zi^{(m)}p_i^\nu-\zj^{(m)}p_j^\nu\Big]\eps^{*}_\mu(\tpij,\lambda')
\eps_\nu(\tpij,\lambda) \right. \\ \nonumber
&& \left. + \frac{\dlpl}{\vijk} \left[ \dlli (1-2\zi)+\dllj (1-2\zj) \right]
\right\} \; ;
\end{eqnarray}
\end{itemize}
where $\dplilj=1-\dlilj$. In each case, the content of the last square
bracket vanishes upon summation over helicities of the emitter
pair. The remaining terms have exactly the same structure as in
\cite{Catani:2002hc} and are only modified by delta's in helicity. The
variables $\zi$, $\zj$ and $\yijk$  are defined as follows
\begin{equation}
\zi = 1-\zj = \frac{p_i p_k}{p_i p_k + p_j p_k} \; ,
\end{equation}
\begin{equation}
\yijk = \frac{p_i p_j}{p_i p_j + p_i p_k + p_j p_k} \; .
\end{equation}
The gluon splitting requires additionally
\begin{equation}
\zi^{(m)} = \zi - \frac{1}{2}(1-\vijk) \; , \quad \quad
\zj^{(m)} = \zj - \frac{1}{2}(1-\vijk) \; ,
\end{equation}
where $\vijk$ is the relative velocity between the emitter pair with
momentum $p_i+p_j$ and the spectator with momentum $p_k$, and can be
expressed as
\begin{equation}
\vijk  = 
\frac{\sqrt{[2\mu_k^2+(1-\mu_i^2-\mu_j^2-\mu_k^2)(1-\yijk)]^2-4\mu_k^2}}
{(1-\mu_i^2-\mu_j^2-\mu_k^2)(1-\yijk)} \; ,
\end{equation}
with
\begin{equation}
\mu_{n}=m_{n}/\sqrt{Q^{2}} \; .
\end{equation}
Similarly, the relative velocity between the emitter with momentum
$\tpij$ and the spectator with momentum $\tpk$ can be written as
\begin{equation}
\tvijk = \frac{\sqrt{\lambda(1,\mu_{ij}^2,\mu_k^2)}}{1-\mu_{ij}^2-\mu_k^2} \; .
\end{equation}
Moreover, the terms proportional to the free parameter $\kappa$
require the introduction of
\begin{equation}
z_\pm(\yijk)  = \frac{2\mu_i^2+(1-\mu_i^2-\mu_j^2-\mu_k^2)\yijk}
{2[\mu_i^2+\mu_j^2+(1-\mu_i^2-\mu_j^2-\mu_k^2)\yijk]} (1\pm\viji\vijk) \; ,
\end{equation}
where $\viji$ is the relative velocity between $p_i+p_j$ and $p_i$
\begin{equation}
\viji  = 
\frac{\sqrt{(1-\mu_i^2-\mu_j^2-\mu_k^2)^2\yijk^2-4\mu_i^2\mu_j^2}}
{(1-\mu_i^2-\mu_j^2-\mu_k^2)\yijk+2\mu_i^2} \; .
\end{equation}
At this point it is important to comment on the freedom in the
modification of the original formulae of \cite{Catani:2002hc}. Since
we only require that these be recovered upon summation over
helicities, arbitrary factors can be introduced as long as they do not
spoil the collinear and/or soft limits. We have used this freedom to
make the terms, which vanish upon helicity summation, resemble those
that do not. Thus, for example, we have introduced the coefficient
$\tvijk/\vijk$ in front of the last square bracket in
Eq.~\ref{eq:ff1}, and similarly $1/\vijk$ in Eq.~\ref{eq:ff3}. A
similar freedom exists in the treatment of the parameter $\kappa$. As
it represents a soft/collinear safe modification, it is unconstrained
in the singular limits, and we are free to distribute it among different
polarizations as we please. Here, we have chosen an even distribution
among the four possible combinations.

%%%%%%%%%%%%%%%%%%%%%%%%%%%%%%%%%%%%%%%%%%%%%%%%%%%%%%%%%%%%%%%%%%%%%%%%%%%%%%%%

\subsection{Final-state emitter and initial-state spectator}

The single dipole contribution is
\begin{eqnarray}
&& \!\!\!\!\!\!
\cD^{a}_{ij}(\dots,\{p_i,\lambda_i\},\dots,\{p_j,\lambda_j\},\dots;
\{p_a,\lambda_a\},\dots) =
-\frac{1}{(p_i+p_j)^2-m_{ij}^2} \frac{1}{\xija} \\ \nonumber
&& \quad \quad \quad \times \sum_{\lambda',\lambda}
\langle\dots,\{\tpij,\lambda'\}, \dots;\{\tpa,\lambda_a\},\dots|
\frac{{\bom T}_a\cdot{\bom T}_{ij}}{{\bom T}_{ij}^2} \bV^{a}_{ij}
|\dots,\{\tpij,\lambda\},\dots;\{\tpa,\lambda_a\},\dots\rangle \; ,
\end{eqnarray}
where
\begin{equation}
\xija  = \frac{p_a p_i+p_a p_j-p_i p_j+\frac{1}{2}(m_{ij}^2-m_i^2-m_j^2)}
{p_a p_i + p_a p_j} \; ,
\end{equation}
and the momentum remapping is given by
\begin{equation}
\tpa^\mu = \xija p_a^\mu \; ,
\end{equation}
\begin{equation}
\tpij^\mu=p_i^\mu+p_j^\mu-(1-\xija)p_a^\mu \; .
\end{equation}
There are three cases to consider
\begin{itemize}
\item
{\boldmath $Q\to Q(p_i,\lambda_i)+g(p_j,\lambda_j)$}, with
$m_i=m_{ij}=m_Q$, $m_j=0$ and
\begin{eqnarray}
\label{eq:fi1}
\langle \lambda' |\bV^{a}_{Qg}| \lambda \rangle &=& 8\pi\alps\CF \\ \nonumber
&\times& \left\{
\frac{\delta_{\lambda\lambda_i}}{1-\zi+(1-\xija)}
-\dplilj \left(
\delta_{\lambda\lambda_i}(1+\zi)+\delta_{\lambda\lambda_j}
\frac{m_Q^2}{p_i p_j} \right)\right. \\ \nonumber 
&& \left. - \left[
(\dlli\dlilj-\dplli\dplilj) \frac{1}{\zi}+\dplilj(\dlli-\dllj)
  \zi\right] \frac{m_Q^2}{2p_ip_j}
\right\} \dlpl \; ;
\end{eqnarray}
\item
{\boldmath $g\to Q(p_i,\lambda_i) + \bar Q(p_j,\lambda_j)$}, with 
$m_i=m_j=m_Q$, $m_{ij}=0$ and
\begin{eqnarray}
\label{eq:fi2}
\langle\lambda'|\bV^{a}_{Q\bar Q}|\lambda\rangle  &=& 
8\pi\alps\TR  \\ \nonumber &\times& 
\left\{
\frac{\dlpl\dplilj}{2} - \frac{2\dplilj}{(p_i+p_j)^2}
\left[\zi p_i^\mu-\zj p_j^\mu\right]\left[\zi p_i^\nu-\zj p_j^\nu\right]
\eps^{*}_\mu(\tpij,\lambda')\eps_\nu(\tpij,\lambda) \right. \\ \nonumber
&& + \dlpl \left. \left[  \dplilj\left(\dlli \zi+\dllj \zj-\frac{1}{2}\right)
\right. \right. \\ \nonumber && \quad \quad \quad \left. \left.
+\left( (\dlli-\dplilj)\frac{1}{\zi}+(\dllj-\dplilj)\frac{1}{\zj}\right)
\frac{m_Q^2}{(p_i+p_j)^2} \right] 
\right\} \; ;
\end{eqnarray}
\item
{\boldmath $g\to g(p_i,\lambda_i) +g(p_j,\lambda_j)$}, with
$m_i=m_j=m_{ij}=0$ and
\begin{eqnarray}
\label{eq:fi3}
\langle\lambda'|\bV^{a}_{gg}|\lambda\rangle  &=&
8\pi\alps\CA \\ \nonumber &\times&
\left\{ \dlpl \left(\frac{\dlli}{1-\zi+(1-\xija)}
+\frac{\dllj}{1-\zj+(1-\xija)}-2\dplilj \right) \right. \\ \nonumber
&& \left. +\frac{\dplilj}{p_i p_j}
\Big[\zi p_i^\mu-\zj p_j^\mu\Big]\Big[\zi p_i^\nu-\zj p_j^\nu\Big]
\eps^{*}_\mu(\tpij,\lambda')\eps_\nu(\tpij,\lambda) \right. \\ \nonumber
&& \left. +\dlpl \left[ \dlli (1-2\zi)+\dllj (1-2\zj) \right]
\right\} \; ;
\end{eqnarray}
\end{itemize}
where
\begin{equation}
\zi = \frac{p_a p_i}{p_a p_i + p_a p_j} \; , \quad \quad
\zj = \frac{p_a p_j}{p_a p_i + p_a p_j} \; .
\end{equation}
As in the previous case, terms in the last square bracket vanish upon
summation over emitter pair helicities.

%%%%%%%%%%%%%%%%%%%%%%%%%%%%%%%%%%%%%%%%%%%%%%%%%%%%%%%%%%%%%%%%%%%%%%%%%%%%%%%%

\subsection{Initial-state emitter and final-state spectator}

The single dipole contribution is
\begin{eqnarray}
&& \!\!\!\!\!\!
\cD^{ai}_{j}(\dots,\{p_i,\lambda_i\},\dots,\{p_j,\lambda_j\},\dots;
\{p_a,\lambda_a\},\dots) =
-\frac{1}{2 p_a p_i} \frac{1}{\xija} \\ \nonumber
&& \quad \quad \quad \times \sum_{\lambda',\lambda}
\langle\dots,\{\tpj,\lambda_j\}, \dots;\{\tpai,\lambda'\},\dots|
\frac{{\bom T}_j\cdot{\bom T}_{ai}}{{\bom T}_{ai}^2} \bV^{ai}_{j}
|\dots,\{\tpj,\lambda_j\},\dots;\{\tpai,\lambda\},\dots\rangle \; ,
\end{eqnarray}
where
\begin{equation}
\xija = \frac{p_a p_i+p_a p_j-p_i p_j}{p_a p_i + p_a p_j} \; ,
\end{equation}
and the momentum remapping is given by
\begin{equation}
\tpj^\mu = p_i^\mu + p_j^\mu - (1-\xija)p_a^\mu \; ,
\end{equation}
\begin{equation}
\tpai^\mu = \xija p_a^\mu \; .
\end{equation}
As implicitly assumed above, we require both emitter masses to vanish,
{\it i.e.} $m_i=m_j=m_{ij}=0$. There are four cases to consider
\begin{itemize}
\item
{\boldmath $q(p_a,\lambda_a) \to g(p_i,\lambda_i) + q$}, with
\begin{eqnarray}
\langle \lambda'|\bV^{qg}_j|\lambda\rangle &=&
8\pi\alps\CF \\ \nonumber &\times&
\left\{
\frac{1}{1-\xija+u_i}-\dplali(1+\xija)\right\} \dlpl \dlla \; ;
\end{eqnarray}
\item
{\boldmath $g(p_a,\lambda_a) \to \bar q(p_i,\lambda_i) + q$}, with 
\begin{eqnarray}
\langle \lambda'|\bV^{g\bar q}_j|\lambda \rangle &=&
8\pi\alps\TR  \\ \nonumber &\times&
\left\{
\dlali(1-2\xija(1-\xija))+[1-2\dlali]\xija^2 \right\} \dlpl \dplli \; ;
\end{eqnarray}
\item
{\boldmath $q(p_a,\lambda_a) \to q(p_i,\lambda_i) + g$}, with
\begin{eqnarray}
\langle\lambda'|\bV^{qq}_j|\lambda\rangle  &=&
8\pi\alps\CF \\ \nonumber &\times& \Big\{
\dlpl\dplla\dlali\xija \\ \nonumber
&& + \dlali \frac{1-\xija}{\xija}\frac{u_i(1-u_i)}{p_i p_j}
\biggl(\frac{p_i^\mu}{u_i}-\frac{p_j^\mu}{1-u_i}\biggr)
\biggl(\frac{p_i^\nu}{u_i}-\frac{p_j^\nu}{1-u_i}\biggr)
\eps_\mu(\tpai,\lambda')\eps^{*}_\nu(\tpai,\lambda) \\ \nonumber
&&
+\dlpl\dlali [ 2\dlla-1 ]
\Big\} \; ;
\end{eqnarray}
\item
{\boldmath $g(p_a,\lambda_a) \to g(p_i,\lambda_i) + g$}, with
\begin{eqnarray}
\langle\lambda'|\bV^{gg}_j|\lambda\rangle  &=&
8\pi\alps\CA \\ \nonumber &\times&
\left\{
\dlpl\biggl[\frac{\dlla}{1-\xija+u_i}+\dplli\left(-1+\xija(1-\xija)\right)
\biggr]\right. \\ \nonumber
&& + \dlali \frac{1-\xija}{\xija}\frac{u_i(1-u_i)}{p_i p_j}
\biggl(\frac{p_i^\mu}{u_i}-\frac{p_j^\mu}{1-u_i}\biggr)
\biggl(\frac{p_i^\nu}{u_i}-\frac{p_j^\nu}{1-u_i}\biggr)
\eps_\mu(\tpai,\lambda')\eps^{*}_\nu(\tpai,\lambda) \\ \nonumber
&& +\dlpl \left[ (\dlla-\dplli)+(\dlali-\dlla) 2 \xija \right]
\Big\} \; ;
\end{eqnarray}
\end{itemize}
where
\begin{equation}
u_i = \frac{p_i p_a}{p_i p_a+p_j p_a} \; .
\end{equation}

%%%%%%%%%%%%%%%%%%%%%%%%%%%%%%%%%%%%%%%%%%%%%%%%%%%%%%%%%%%%%%%%%%%%%%%%%%%%%%%%

\subsection{Initial-state emitter and initial-state spectator}

The single dipole contribution is
\begin{eqnarray}
&& \!\!\!\!\!\!\!\!\!\!
\cD^{ai,b}(\dots,\{p_i,\lambda_i\},\dots;\{p_a,\lambda_a\},\{p_b,\lambda_b\}) =
-\frac{1}{2 p_a p_i} \frac{1}{\xiab} \\ \nonumber
&& \quad \quad \quad \quad \quad \quad \quad \quad \quad \times
\sum_{\lambda',\lambda}
\langle{\widetilde \dots};\{\tpai,\lambda'\},\{p_b,\lambda_b\}|
\frac{{\bom T}_b\cdot{\bom T}_{ai}}{{\bom T}_{ai}^2} \bV^{ai,b}
|{\widetilde \dots};\{\tpai,\lambda\},\{p_b,\lambda_b\}\rangle \; ,
\end{eqnarray}
where
\begin{equation}
\xiab = \frac{p_a p_b - p_i p_a - p_i p_b}{p_a p_b} \; ,
\end{equation}
and the momentum remapping is given by
\begin{equation}
{\widetilde p}_{ai}^\mu = x_{i,ab} \,p_a^\mu  \; ,
\end{equation}
\begin{eqnarray}
{\widetilde k_j}^{\mu} = k_j^{\mu} - \frac{2 k_j \cdot (K+{\widetilde K})}
{(K+{\widetilde K})^2} \;(K+{\widetilde K})^{\mu} + \frac{2 k_j \cdot K}
{K^2} \;{\widetilde K}^{\mu} \; ,
\end{eqnarray}
where the index $j$ runs over all final states and the momenta
$K^{\mu}$ and ${\widetilde K}^\mu$ are defined by
\begin{equation}
\begin{array}{rcl}
K^{\mu} &=& p_a^\mu + p_b^\mu - p_i^\mu \;\;, \\
{\widetilde K}^{\mu} &=& {\widetilde p_{ai}}^\mu + p_b^\mu \;\;.
\end{array}
\end{equation}
We require again both emitter masses to vanish, {\it i.e.}
$m_i=m_j=m_{ij}=0$. There are four cases to consider
\begin{itemize}
\item
{\boldmath $q(p_a,\lambda_a) \to g(p_i,\lambda_i) + q$}, with
\begin{eqnarray}
\langle \lambda'|\bV^{qg,b}|\lambda\rangle &=&
8\pi\alps\CF \\ \nonumber &\times&
\left\{
\frac{1}{1-\xiab}-\dplali(1+\xiab)\right\} \dlpl \dlla \; ;
\end{eqnarray}
\item
{\boldmath $g(p_a,\lambda_a) \to \bar q(p_i,\lambda_i) + q$}, with 
\begin{eqnarray}
\langle \lambda'|\bV^{g\bar q,b}|\lambda \rangle &=&
8\pi\alps\TR  \\ \nonumber &\times&
\left\{
\dlali(1-2\xiab(1-\xiab))+[1-2\dlali]\xiab^2 \right\} \dlpl \dplli \; ;
\end{eqnarray}
\item
{\boldmath $q(p_a,\lambda_a) \to q(p_i,\lambda_i) + g$}, with
\begin{eqnarray}
\langle\lambda'|\bV^{qq,b}|\lambda\rangle  &=&
8\pi\alps\CF \\ \nonumber &\times& \Big\{
\dlpl\dplla\dlali\xiab \\ \nonumber
&& + \dlali \frac{1-\xiab}{\xiab} \;
\frac{p_a \cdot p_b}{p_i \cdot p_a \;p_i \cdot p_b} \,
\left( p_i^{\mu} - \frac{p_ip_a}{p_bp_a} p_b^{\mu} \right)
\left( p_i^{\nu} - \frac{p_ip_a}{p_bp_a} p_b^{\nu} \right) \\ \nonumber
&& \times \eps_\mu(\tpai,\lambda')\eps^{*}_\nu(\tpai,\lambda)
+\dlpl\dlali [ 2\dlla-1 ]
\Big\} \; ;
\end{eqnarray}
\item
{\boldmath $g(p_a,\lambda_a) \to g(p_i,\lambda_i) + g$}, with
\begin{eqnarray}
\langle\lambda'|\bV^{gg,b}|\lambda\rangle  &=&
8\pi\alps\CA \\ \nonumber &\times&
\left\{
\dlpl\biggl[\frac{\dlla}{1-\xiab}+\dplli\left(-1+\xiab(1-\xiab)\right)
\biggr]\right. \\ \nonumber
&& + \dlali \frac{1-\xiab}{\xiab} \;
\frac{p_a \cdot p_b}{p_i \cdot p_a \;p_i \cdot p_b} \,
\left( p_i^{\mu} - \frac{p_ip_a}{p_bp_a} p_b^{\mu} \right)
\left( p_i^{\nu} - \frac{p_ip_a}{p_bp_a} p_b^{\nu} \right) \\ \nonumber
&& \times \eps_\mu(\tpai,\lambda')\eps^{*}_\nu(\tpai,\lambda)
+\dlpl\left[ (\dlla-\dplli)+(\dlali-\dlla) 2 \xiab \right]
\Big\} \; .
\end{eqnarray}
\end{itemize}

%%%%%%%%%%%%%%%%%%%%%%%%%%%%%%%%%%%%%%%%%%%%%%%%%%%%%%%%%%%%%%%%%%%%%%%%%%%%%%%%

\section{Implementation and example results}
\label{sec:implementation}

As part of the present study, we have implemented the complete
formalism given in the previous section within the framework of
\textsc{Helac-Phegas}. The software can be obtained from the 
\textsc{Helac-Phegas} web page \cite{webdip}. 
The main features can be summarized as follows
\begin{enumerate}

\item {\bf Arbitrary processes}

  We use the matrix element generator of \textsc{Helac-Phegas}, which
  means that any process, which can be calculated by this generator is
  also accessible to the dipole subtraction software. The only
  limitation is given by the models implemented. Currently, the full
  Standard Model, both electroweak and QCD, is available. Let us
  stress that the subtraction is only applied to partons, and the
  software is therefore devised for NLO QCD corrections. The
  observables are specified by user defined jet functions and
  histogramming routines. The present implementation contains a built
  in $k_T$ jet algorithm.

\item {\bf Massive and massless external states}

  We have implemented the formulae as given in the previous
  section. This allows us to treat massive and massless partons on the
  same footing. As noted in the previous section, we assume the
  initial state partons to be massless.

\item {\bf Helicity sampling for partons}

  The integration over the phase space is done in three stages. At
  first, the phase space is optimized using multi-channel methods. In
  this phase, the user can either evaluate the subtracted matrix
  elements with full summation,  or use a flat helicity Monte Carlo,
  where all helicity configurations occur with the same
  frequency. Since this part is only used to determine the weights of
  the different channels we recommend to run in the latter mode, which
  is fast and gives exactly the same results as full summation.  In a
  second stage, the phase space measure is fixed and the helicity
  configuration sampling weights are determined by evaluating the
  subtracted matrix element for all helicity configurations. The
  optimization consists of a standard minimization of the variance
  \cite{Kleiss:1994qy}, but is as slow as full summation (in fact, this
  {\it is} full summation). We recommend to run this phase on a
  few hundreds up to a few thousand points. The optimum depends
  strongly on the number of helicity configurations. Finally, in
  the last stage, the matrix element is evaluated with helicity
  sampling, which updates itself at user specified intervals in
  the number of accepted events. These intervals should be long
  enough, to allow the channels which have a low weight to
  accumulate enough events. We recommend a number of the order of
  ten thousand. All the parameters can be changed by the
  user. Obviously, one can make a complete run using full
  helicity summation of flat Monte Carlo. As with any simulation,
  some experimentation is needed to obtain best results.

\item {\bf Random polarizations for non-partons}

  Non-partons can be treated in two different ways as far as their
  polarization is concerned. The user can either require random phases
  as described in Section~\ref{sec:polarization}, or treat non-partons
  on the same footing as partons. We recommend to use random
  polarizations.

\item {\bf Restrictions on the subtraction phase space}

  We have implemented a restriction on the phase space of the
  subtraction as proposed in \cite{Nagy:2003tz}. This amounts to only
  including dipole subtraction terms, which satisfy the following
  criteria
  \begin{enumerate}
  \item final-final dipoles
    
    \begin{equation}
      \yijk < \alpha_{max}^{FF} \; ;
    \end{equation}

  \item final-initial dipoles
    
    \begin{equation}
      1-\xija < \alpha_{max}^{FI} \; ;
    \end{equation}

  \item initial-final dipoles
    
    \begin{equation}
      u_i < \alpha_{max}^{IF} \; ;
    \end{equation}

  \item initial-initial dipoles
    
    \begin{equation}
      {\tilde v}_i \equiv \frac{p_a p_i}{p_a p_b} <
      \alpha_{max}^{II} \; ;
    \end{equation}
    
  \end{enumerate}
  where the occurring variables have been defined in the previous
  section. The four $\alpha_{max}$ parameters can be freely chosen by the
  user. A value of $1$ amounts to no restriction. Best results are
  obviously obtained with small $\alpha_{max}$, since the matrix
  element for this setting behaves largely as the real emission alone
  and the phase space optimizer has been designed to emulate such
  behavior. Besides the phase space restriction, we have also
  included a technical cut. Using the above formulae one can evaluate
  a minimal value of the four parameters given on the left hand
  side for all dipoles. Let us denote it by $\alpha_{min}$. We reject
  a phase space point completely if
  \begin{equation}
    \alpha_{min} < \alpha_{cut} \; .
  \end{equation}
  The value of $\alpha_{cut}$ should be specified by the user. Let us
  stress that the technical cut is necessary to avoid numerical
  instabilities in the cancellation between real emission and the
  dipoles. However, there are also cutoffs in the phase space
  generator, which are needed to avoid numerical instabilities in the
  generator itself.

\item {\bf Phase space integration}

  The integration is completely controlled by
  \textsc{Phegas} \cite{Papadopoulos:2000tt}, which is a multi-channel 
  phase space generator based on Feynman graphs. The only modification, 
  which was necessary was the simultaneous treatment of both positive 
  and negative weights. In fact, the user can specify whether only positive,
  negative or both weights should be included. This does not modify
  the histograms, however, which always contain the full result.

\end{enumerate}

We should mention that in the early stages of the development, we have
adapted the kinematics remapping routine from \textsc{Mcfm}
\cite{mcfm}. Although, a minute part of the final code, it was
important for debugging to be able to rely on the correctness of the
kinematics.

This software has been tested against \textsc{MadDipole}
\cite{maddipole}. We would like, however, to point out that instead of
comparing arbitrary processes, we concentrated on the individual parts
of the code. In fact, the only sensitive code, which cannot be tested
against the official version of \textsc{Helac-Phegas} is contained in the
kinematics remapping, the color correlators and the dipole subtraction
formulae themselves. In consequence, to ensure the correctness of the
implementation it was sufficient to compare the latter for all
independent cases. These are summarized in Tab.~\ref{tab:check}. Of
course, we could only compare the dipole subtraction terms after
summation over helicities. The expressions for independent
helicity configurations have been tested by checking for cancellation
in the appropriate limits.
%%====================================================     
\begin{table}[h!]
{\footnotesize
\begin{center}
${\cal{E}}_{0}$ - massless emitter, ${\cal{S}}_{0}$ - massless spectator, 
${\cal{E}}_{M}$ - massive emitter, ${\cal{S}}_{M}$ - massive spectator, 
${\cal{E}}_{I}$ - initial state emitter, ${\cal{E}}_{F}$ - final state emitter,
${\cal{S}}_{I}$ - initial state spectator, ${\cal{S}}_{F}$ - final state 
spectator, $\checkmark$ - check, $\blacksquare$ - does not occur.
\end{center}
\begin{minipage}
[b]{0.45\linewidth}\centering
\begin{tabular}{c|c|c|c|c}
%\hline
%\hline
&&&&\\
 & ${\cal{E}}_{0}/{\cal{S}}_{0}$& 
${\cal{E}}_{0}/{\cal{S}}_{M}$& 
${\cal{E}}_{M}/{\cal{S}}_{0}$&
${\cal{E}}_{M}/{\cal{S}}_{M}$\\
&&&&\\
\hline
&&&&\\
${\cal{E}}_{I}$/${\cal{S}}_{I}$&&&&\\
&&&&\\
$g \to gg$ & 
$\checkmark$   & $\blacksquare$ &$\blacksquare$& $\blacksquare$\\
$g \to qq $  &$\checkmark$&$\blacksquare$&$\blacksquare$&$\blacksquare$\\
$q \to qg$ &$\checkmark$&$\blacksquare$&$\blacksquare$& $\blacksquare$\\
$q \to gq$ &$\checkmark$&$\blacksquare$&$\blacksquare$&$\blacksquare$\\
 & & &&\\
\hline
 & & &&\\
${\cal{E}}_{I}/{\cal{S}}_{F}$&&&&\\
& & &&\\
$g \to gg$ &$\checkmark$&$\checkmark$&$\blacksquare$&$\blacksquare$ \\
$g \to qq $  &$\checkmark$&$\checkmark$&$\blacksquare$&$\blacksquare$\\
$q \to qg$ &$\checkmark$&$\checkmark$&$\blacksquare$&$\blacksquare$ \\
$q \to gq$ &$\checkmark$&$\checkmark$&$\blacksquare$&$\blacksquare$\\
 & & &&\\
%\hline
%\hline
\end{tabular}
%\end{table}

%\begin{table}
\end{minipage}
\hspace{1cm}
\begin{minipage}[b]{0.45\linewidth}\centering
\begin{tabular}{c|c|c|c|c}
%\hline
%\hline
&&&&\\
 & ${\cal{E}}_{0}/{\cal{S}}_{0}$& 
${\cal{E}}_{0}/{\cal{S}}_{M}$& 
${\cal{E}}_{M}/{\cal{S}}_{0}$&
${\cal{E}}_{M}/{\cal{S}}_{M}$\\
&&&&\\
\hline
& & &&\\
${\cal{E}}_{F}/{\cal{S}}_{I}$&&&& \\
& & &&\\
$g \to gg$ & $\checkmark$ &$\blacksquare$&$\blacksquare$& $\blacksquare$\\
$g \to qq $  & $\checkmark$&$\blacksquare$&$\checkmark$&$\blacksquare$\\
$q \to qg$ & $\checkmark$&$\blacksquare$&$\checkmark$& $\blacksquare$\\
$q \to gq$ & $\checkmark$&$\blacksquare$&$\checkmark$&$\blacksquare$\\
 & & &&\\
\hline
&&&& \\
${\cal{E}}_{F}/{\cal{S}}_{F}$&&&& \\
& &&&\\
$g \to gg$ &$\checkmark$&$\checkmark$&$\blacksquare$& $\blacksquare$\\
$g \to qq $  &$\checkmark$&$\checkmark$&$\checkmark$&$\checkmark$\\
$q \to qg$ &$\checkmark$&$\checkmark$&$\checkmark$&$\checkmark$ \\
$q \to gq$ &$\checkmark$&$\checkmark$&$\checkmark$&$\checkmark$\\
 & & &&\\
%\hline
%\hline
\end{tabular}
\end{minipage}
\vspace{0.4cm}
\caption{\label{tab:check} \it Independent dipole splitting formulae,
  which need to be tested in order to ensure the correctness of the
  code. In the splitting description, {\it e.g.} $g \rightarrow gg$,
  the left hand side particle always denotes the virtual state.}}
\end{table}
%%====================================================     
\begin{table}[!t]
{\footnotesize
\begin{center}
\begin{tabular}{c|c|c|c}
&& &\\
\textsc{Process} &  \textsc{Real Emission + Dipoles} 
&  \textsc{Real Emission} &  \textsc{Nr Of Dipoles}\\
& [msec] & [msec] &\\
&&&\\
\hline
&&&\\
$gg \rightarrow ggg$   & 3.8 &  1.0 &   27\\
$gg \rightarrow gggg$  &  8.5  &  2.6 &   56 \\
$gg \rightarrow ggggg$  &  300  &  42 &  100 \\
&&&\\
\hline
&&&\\
$u\bar{d} \rightarrow W^{+}gggg$    &9.3 &   2.4&    56\\
&&&\\
\hline
&&&\\
$gg \rightarrow t\bar{t}b\bar{b}g$  &  12 &   2.9&    55\\
&&&\\
\end{tabular}
\vspace{0.4cm}
\caption{\label{tab:time}
  \it The CPU time needed to evaluate the real emission matrix
  element together with all of the dipole subtraction terms per
  phase-space point (this corresponds to $\alpha_{max}=1$). All
  numbers have been obtained on an Intel 2.53 GHz Core 2 Duo processor
  with the Intel Fortran compiler using the {\it -fast} option. }
\end{center}}
\end{table}
%%====================================================     
\begin{figure}[t]
\begin{center}
\includegraphics[width=0.45\textwidth]{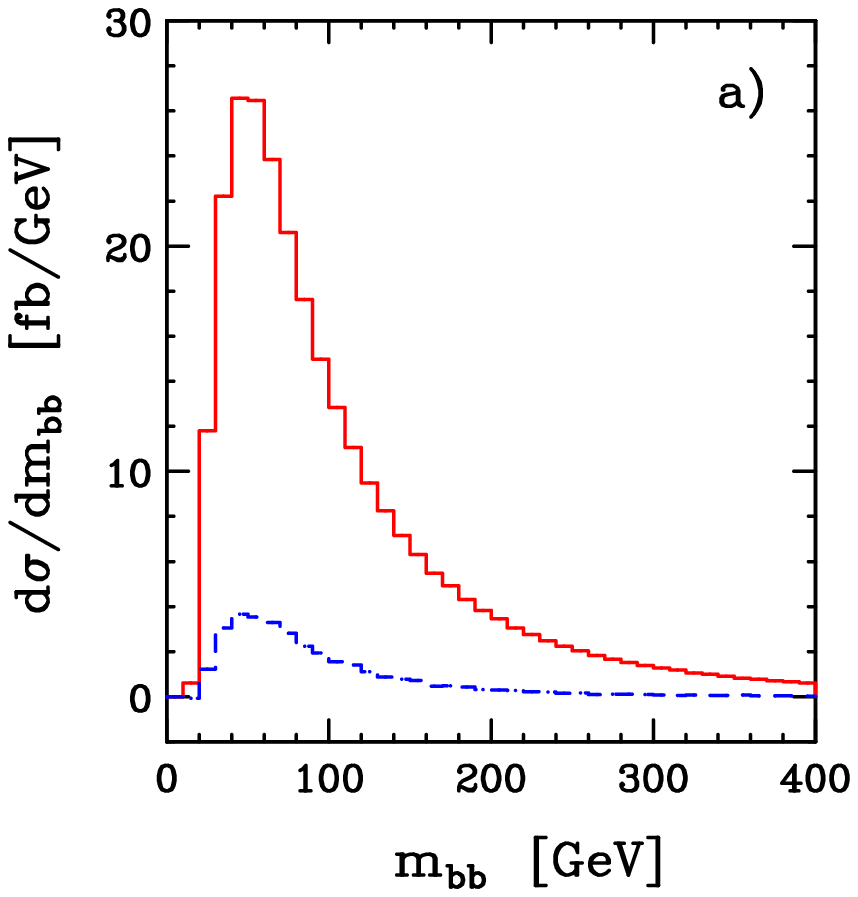}
\includegraphics[width=0.45\textwidth]{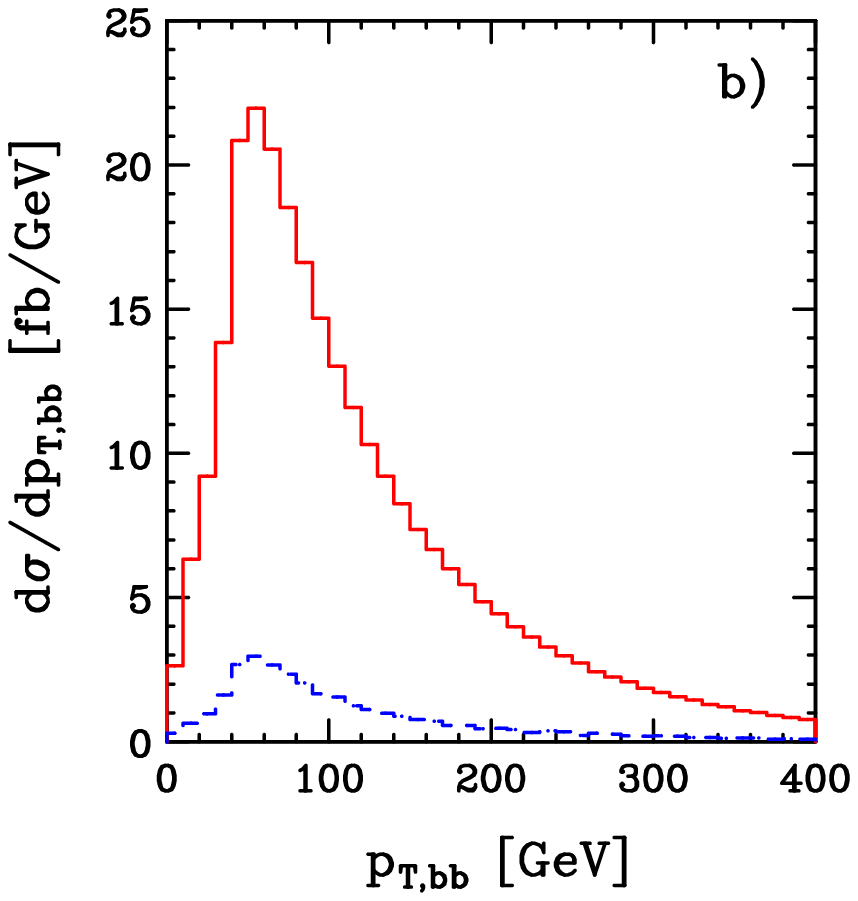}
\includegraphics[width=0.45\textwidth]{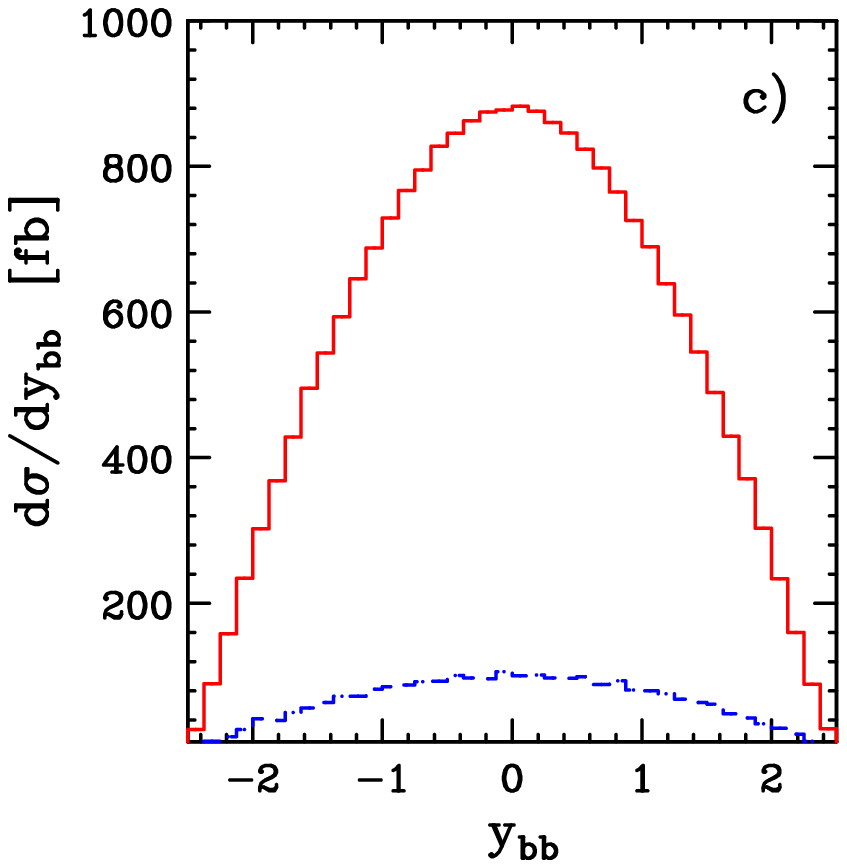}
\end{center}
\vspace{-0.2cm}
\caption{\it \label{fig:distros} Distribution of the invariant mass
  $m_{b\bar{b}}$ of the bottom-anti-bottom pair (a), distribution in
  the transverse momentum  $p_{T_{b\bar{b}}}$  of the
  bottom-anti-bottom pair (b) and distribution in the rapidity
  $y_{b\bar{b}}$ of the bottom-anti-bottom pair (c) for
  $pp(gg)\rightarrow t\bar{t}b\bar{b}g + X$ at the LHC for
   two different values of $\alpha_{max}$: 0.01 (red solid curve) and 
   1 (blue dashed curve).}
%. Distribution of the invariant mass
%  $m_{b\bar{b}}$ of the  bottom-anti-bottom pair (d) for three
%  different values of $\alpha_{max}$: 0.01  (black curve), 0.1 (red
%  curve) and 1.0 (blue curve).}
\end{figure}
%%====================================================   

Additionally, in Tab.~\ref{tab:time} we have presented the measured
time needed to evaluate the real emission matrix element and 
the subtraction terms. We note that the
inclusion of the full set of subtraction terms slows down the
computation by a factor of about three to four in most cases. However,
since there is a restriction in the subtraction phase space, we expect
(and indeed observe in practice) that the true additional cost of the
subtraction does not exceed the cost of the real emission
itself. Therefore, an improvement in the speed of the calculation will
not be obtained by replacing the dipole formalism by another
subtraction. In fact, a further substantial speed up can be obtained
by turning to Monte Carlo summation over color configurations. What is
the source of the high cost of the evaluation of the subtracted real
radiation in comparison to leading order real radiation? It lies in
the phase space integration. The integrand has now a much more
complicated behavior, and therefore multi-channel optimization based
on Feynman graphs does not lead to such a drastic improvement of the
convergence.

As a final demonstration of the power of our implementation, we have
performed a simulation of the $gg \rightarrow t {\bar t} b {\bar b} g$
subprocess, which is part of the complete NLO calculation of $pp
\rightarrow t {\bar t} b {\bar b} + X$. We have used the same set up
as \cite{Bredenstein:2008zb}.  In particular, we have taken
$\sqrt{s}=14$ TeV as center of mass energy and a top quark mass of
$m_t=172.6$ GeV. The b quark has been kept massless. In order to
obtain jets, we have used the $k_T$ algorithm
\cite{Catani:1992zp,Catani:1993hr,Ellis:1993tq}  with jet
recombination of partons with a pseudo-rapidity of $|\eta| < 5$ with
$\Delta R < 0.8$. Additionally, we required the b jets to satisfy
$p_{T,b} > 20$ GeV, $|y_b| < 2.5$. The phase space of the top quarks
has not been restricted. The non-perturbative input of our computation
has been given by the CTEQ6M PDFs
\cite{Pumplin:2002vw,Stump:2003yu}. Fig.~\ref{fig:distros} (a-c)
contains the invariant mass, transverse momentum and rapidity
distributions  of the $b\bar{b}$ pair for $\alpha_{max}=0.01$ and 
$\alpha_{max}=1$ (all parameters set to the same value).  The impact 
of the variation of $\alpha_{max}$ is cleary visible. In fact, for both values,
the shape resembles that of the leading order result, besides the
normalization. On the other hand, for  
larger values of $\alpha_{max}$ a relatively small negative dip for low 
invariant masses can be observed in the $m_{b\bar{b}}$ distribution. The latter 
is due to the strong enhacement of the dipole contributions for low 
$m_{b\bar{b}}$.

%%%%%%%%%%%%%%%%%%%%%%%%%%%%%%%%%%%%%%%%%%%%%%%%%%%%%%%%%%%%%%%%%%%%%%%%%%%%%%%%

\section{Conclusions}
\label{sec:conclusions}

We have presented an extension of the dipole subtraction formalism to
arbitrary helicity eigenstates. Not only did we provide appropriate
formulae, but also a public implementation, which contains both the
subtracted real radiation and integrated dipoles. The results of our tests
show that with this software, calculations, which are of current
phenomenological interest, can be performed fully automatically within
tractable time. In the nearest future, we will apply this technology
to the processes from the ``NLO wishlist'' \cite{Bern:2008ef} using the
automated 1-loop extension of \textsc{Helac} 
\cite{vanHameren:2009dr} for the virtual corrections.

%%%%%%%%%%%%%%%%%%%%%%%%%%%%%%%%%%%%%%%%%%%%%%%%%%%%%%%%%%%%%%%%%%%%%%%%%%%%%%%%

\section*{Acknowledgments}

The work of M.C. was supported in part by the ToK Program 
  ALGOTOOLS (MTKD-CD-2004-014319) and by the Heisenberg Programme of
  the Deutsche Forschungsgemeinschaft. M.W. was funded in part by the
RTN European Programme MRTN-CT-2006-035505 HEPTOOLS - Tools and
Precision Calculations for Physics Discoveries at Colliders and by the
Initiative and Networking Fund of the Helmholtz Association, contract
HA-101 ("Physics at the Terascale").

%%%%%%%%%%%%%%%%%%%%%%%%%%%%%%%%%%%%%%%%%%%%%%%%%%%%%%%%%%%%%%%%%%%%%%%%%%%%%%%%

%\bibliographystyle{utphys_spires}
%\bibliography{DIPOLES}

\providecommand{\href}[2]{#2}

\end{document}